\documentclass[prl,aps,twocolumn,superscriptaddress,showpacs]{revtex4}
\usepackage[dvips]{graphicx}
\usepackage{amsmath}
\usepackage{amssymb}

\begin{document}

\title{Critical Temperature and Thermodynamics of Attractive
Fermions at Unitarity}
\author{Evgeni Burovski}
\affiliation{Department of Physics, University of Massachusetts,
Amherst, MA 01003}
\author{Nikolay Prokof'ev}
\affiliation{Department of Physics, University of Massachusetts,
Amherst, MA 01003}%
\affiliation{Russian Research Center ``Kurchatov Institute'',
123182 Moscow, Russia}
\affiliation{Dipartimento di Fisic{\'a}, Universita di Trento and BEC-INFM, I-38050 Povo, Italy}
\author{Boris Svistunov}
\affiliation{Department of Physics, University of Massachusetts,
Amherst, MA 01003}%
\affiliation{Russian Research Center ``Kurchatov Institute'',
123182 Moscow, Russia}
\author{Matthias Troyer}
\affiliation{Institut f{\"u}r theoretische Physik, ETH
Z\"urich, CH-8093 Z{\"u}rich, Switzerland}

\begin{abstract}
The unitarity regime of the BCS-BEC crossover can be realized by
diluting a system of two-component lattice fermions with an
on-site attractive interaction. We perform a systematic-error-free
finite-temperature simulations of this system by diagrammatic
determinant Monte Carlo. The critical temperature in units of
Fermi energy is found to be $T_c / \varepsilon_F = 0.152(7)$. We
also report the behavior of the thermodynamic functions, and
discuss the issues of thermometry of ultracold Fermi gases.
\end{abstract}

\pacs{03.75.Ss,05.10.Ln,71.10.Fd}

\maketitle


The unitarity limit is commonly referred to as the limit of a
diverging scattering length $a \to \infty$, and an effective
range of the interaction $r_e \to 0$. A Fermi gas in this limit
attains  universality: at low enough
temperature the only relevant length scale is given by the
density, $n$, since the divergent scattering
length drops out completely and the system's properties are independent
of the interaction details.
The unitarity limit is approximately realized in the inner crust
of the neutron stars, where the neutron-neutron scattering length
is nearly an order of magnitude larger than the mean interparticle
separation \cite{neutron-star}.
Unitarity conditions can also be achieved with cold
trapped atom gases using the Feshbach resonance technique,
\textit{i.e.} tuning the scattering length to infinity using the
magnetic field. In recent years these systems have been
extensively studied experimentally
\cite{BEC-BCS-expt,Thomas_expt}.

In the limit of $\xi = 1/na^3 \to +\infty$ the fermions pair into
bosonic molecules and form a Bose-Einstein condensate (BEC).  In
the opposite limit $\xi \to -\infty$ one recovers the
Bardeen--Cooper--Schrieffer (BCS) limit. The unitarity limit $\xi
\to 0$ separates these two extremes. In all these cases, a gas
undergoes a superfluid (SF) phase transition at some temperature,
which depends on~$\xi$.

The early analytical treatments of the unitary Fermi gas have been
based on the extension of the BCS-type many-body wave function
\cite{classiki}. Most of the subsequent elaborations are also of
mean-field type (with or without fluctuations)
\cite{mf,Ohashi-Griffin,Holland-Timmermans,LiuHu,Perali,
Haussmann94}. The accuracy and reliability of such approximations
is nevertheless questionable given the strongly interacting nature
of the unitarity regime, and the results differ by nearly an order
of magnitude.

Monte Carlo (MC) simulations of Fermi systems are, in general,
severely hindered by a sign problem \cite{Binder-Landau-book}.
Fortunately for fermions with attractive contact interaction the
sign problem can be avoided \cite{Hirsch,Rubtsov,Kaplan}. The
ground state of a unitary Fermi gas has been studied within a
fixed-node MC framework \cite{Giorgini-Carlson}, the systematic
errors of which depend on the quality of a guess of the nodal
structure of a many-body fermion configuration. Despite a number
of calculations at finite temperatures
\cite{Bulgac,Wingate,Lee-Schaefer}, a reliable estimate of the
critical temperature is lacking. The purpose of this Letter is to
provide accurate results for the critical temperature and
thermodynamic functions of a three-dimensional (3D) unitary Fermi
gas using a novel determinant diagrammatic MC method free of
systematic errors.


Consider an attractive Hubbard model (AHM) defined by the
Hamiltonian $H = H_0 + H_1$, with
\begin{equation}
H_0 = \sum_{\mathbf{k}\sigma} \left( \epsilon_\mathbf{k} - \mu
\right)
c^{\dagger}_{\mathbf{k}\sigma}c_{\mathbf{k}\sigma},~~~~%
H_1 =- U \sum_\mathbf{x} n_{\mathbf{x}\uparrow}
n_{\mathbf{x}\downarrow},
\label{AHM}
\end{equation}
where $c^{\dagger}_{\mathbf{k}\sigma}$ is a fermion creation
operator, $n_{\mathbf{x}\sigma} = c^{\dagger}_{\mathbf{x}\sigma}
c_{\mathbf{x}\sigma}$, $\sigma = \uparrow, \downarrow$ is the spin
index, $\mathbf{x}$ enumerates sites of a 3D simple cubic lattice
with periodic boundary conditions, the quasimomentum $\mathbf{k}$
spans the corresponding Brillouin zone, $\epsilon_\mathbf{k}=-2t
\sum_{\alpha=1}^{3}\cos k_\alpha$ is the  tight-binding spectrum,
$t=1$ is the the hopping amplitude, $\mu$ stands for the chemical
potential, $U>0$ is the on-site attraction, and we have set the
lattice spacing to unity.

By solving the two-body problem of the model (\ref{AHM}) one finds
that the scattering length diverges at $U_c = \left( L^{-3}
\sum_{\mathbf{k} \in \mathrm{BZ}} 1/2\epsilon_k \right)^{-1}
\approx 7.915 t$. We use this value of $U$ throughout. Since we
are ultimately interested in the continuum rather than lattice
results, we study the low density limit $\nu\rightarrow 0$, where
$0 \leqslant \nu \leqslant 2$ is the filling fraction. We define
Fermi momentum as $k_F = (3\pi^2 \nu)^{1/3}$ and Fermi energy
$\varepsilon_F = k_F^2$, as those of a continuum gas with the same
effective mass and number density $n=\nu$.

\begin{figure}
\includegraphics[width=0.99\columnwidth,keepaspectratio=true]{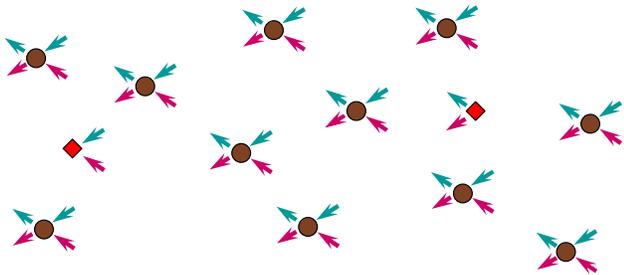}
\caption{A sketch of a vertex configuration for the correlation
function. Brown dots are the four-point vertices, with the
incoming and outgoing lines shown. Red diamonds represent the
two-point vertices corresponding to $P(\mathbf{x},\tau)$ and
$P^\dagger(\mathbf{x}',\tau')$. See the text for discussion. }
\label{fig:conf}
\end{figure}

We simulate the model (\ref{AHM}) by diagrammatic determinant
MC, discussed in detail in Refs.\ \cite{Rubtsov,we}.
One starts by expanding $\exp(-\beta H)$
in the interaction representation in powers of $H_1$. The
resulting Feynmann diagrams consist of four-point vertices
representing the Hubbard interaction, connected by free
single-particle propagators. The sum over all possible ways of
connecting vertices with propagators, in the $n$-th order diagram
is represented by a vertex configuration $\mathcal{S}_n =
\{(\mathbf{x}_j,\tau_j),j=1,\dots,n)\}$, where $\tau$ is the
imaginary time, see Fig.\ \ref{fig:conf} . In case of equal number
of spin-up and spin-down particles, the differential weight of a
configuration is positive definite:
\begin{equation}
d\mathcal{P}(\mathcal{S}_n) = U^n
|\det\mathbf{A}(\mathcal{S}_n)|^2 \prod_{j=1}^n d\tau_j,
\label{MC_weight}
\end{equation}
where $\mathbf{A}(\mathcal{S}_n)$ is an $n \times n$ matrix built
on single-particle propagators:
 $A_{ij} = G^{(0)}(\mathbf{x}_i - \mathbf{x}_j, \tau_i - \tau_j)$.

The configuration space is sampled with worm-type \cite{worm}
updating scheme  \cite{inpreparation}, based on the two-particle
correlation function
\begin{equation}
G_2(\mathbf{x},\tau; \mathbf{x}',\tau') = %
\langle \mathcal{T}_\tau P(\mathbf{x},\tau)
P^\dagger(\mathbf{x}',\tau') \rangle \mathcal{N}^{-2},
\label{corr}
\end{equation}
where $\mathcal{T}_\tau$ is the $\tau-$ordering,
$P(\mathbf{x},\tau) = c_{\mathbf{x}\uparrow}(\tau)
c_{\mathbf{x}\downarrow}(\tau)$ is the pair annihilation operator,
a normalization factor $\mathcal{N}=\beta L^3$ is introduced for
future convenience ($\beta$ is an inverse temperature), and
$\langle \cdots \rangle$ is the thermal average. The non-zero
asymptotic value of $\iint d\tau d\tau' G_2(\mathbf{x},\tau;
\mathbf{x}',\tau')$ as $|\mathbf{x}-\mathbf{x}'| \to \infty$ is
proportional to the condensate density.

Fig.\ \ref{fig:conf} shows a sketch of a vertex configuration,
which features a pair of two-point vertices associated with
$P(\mathbf{x},\tau)$ and $P(\mathbf{x}',\tau')$.

%


The typical number of vertices in a configuration scales with the
system volume as $ M \propto  \beta U L^3$.
Thus the Metropolis acceptance ratios for the updates
involve the ratio of \emph{macroscopically large}
determinants $\det \mathbf{A}(\mathcal{S}'_{n'})/\det
\mathbf{A}(\mathcal{S}_{n})$ with $n'=n$ or $n\pm1$.
Since we only need ratios of determinants, fast-update formulas
\cite{Rubtsov} can be used to reduce the
computational complexity of an update from $M^3$ down to
$M^2$.


We validate our method by comparing results against the exact
diagonalization data for a $4\times 4$ cluster \cite{Husslein},
and simulations of a critical temperature at quarter filling
\cite{Zotos,DCA}. In both cases we find agreement within a few
percent accuracy.

We work in the grand canonical ensemble at fixed
$(L,~T,~\mu )$.
Extracting the unitarity limit critical temperature, $T_c$,
for the continuum gas from the lattice simulation is a two-stage
process: first, we study the  thermodynamic limit
$L \to \infty$ to obtain $T_c(\nu )$ at a given $\nu$  
,and then extrapolate to the continuum limit  $\nu \to 0$.

\begin{figure}
\includegraphics[width=0.99\columnwidth,keepaspectratio=true]{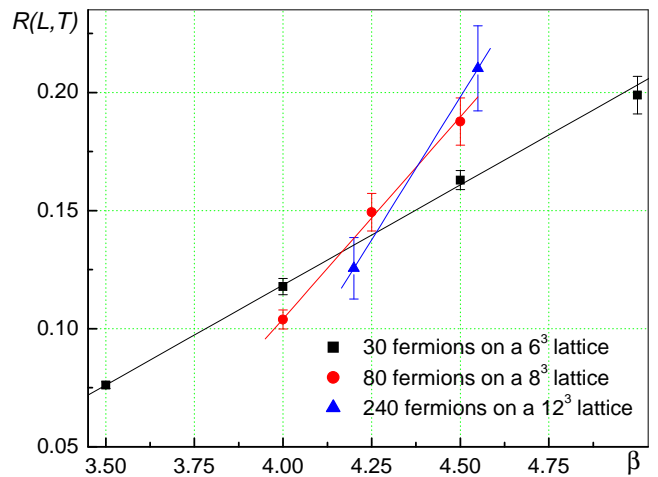}
\caption{A typical crossing of the $R(L,T)$ curves.  The errorbars
are $2\sigma$, and solid lines are the linear fits to the MC
points. The data are for the $\mu=-5.2t$, thus
$\nu(T=T_c,\L\to\infty) = 0.148(1)$.}
 \label{fig:crossing}
\end{figure}

The first task is performed as follows: we simulate a series of
system sizes $L_1 > L_2 > \dots$ at various temperatures. At the
critical point the correlation function (\ref{corr}) decays as a
power-law at large distances: $G_2(\mathbf{x},\tau;
\mathbf{x}',\tau') \propto 1 / |\mathbf{x} -
\mathbf{x}'|^{1+\eta}$, where $\eta$ is an anomalous dimension.
Since the transition is expected to belong to the $U(1)$
universality class, we use $\eta = 0.038$ \cite{Fisher}. Hence, if
one sums and rescales the correlation function (\ref{corr})
according to
\begin{equation}
R(L,T) = L^{1+\eta} \sum_{\mathbf{x}\mathbf{x}'} \int_0^{\beta}
d\tau \int_0^{\beta}d\tau' G_2(\mathbf{x},\tau; \mathbf{x}',\tau'),
\label{rescaled}
\end{equation}
the intersection of the curves $R(L_i,T)$ and $R(L_j,T)$, shown in
Fig.~\ref{fig:crossing}, gives a size-dependent estimate
$T_{L_i,L_j}(\mu)$  for the critical temperature $T_c(\mu)$
\cite{Binder}. As $L \to \infty$, the series of $T_{L_i,L_j}(\mu)$
converges to $T_c(\mu)$ and one can analyze it using corrections
to scaling, to extract its limiting value
\cite{Binder-Landau-book}.
Likewise, a linear fit of a size-dependent estimate for the filling
factor $\nu(L; \mu)$ versus $1/L$ yields the thermodynamic limit
filling factor $\nu(\mu)$.

The next step is to repeat the procedure for a sequence of $\mu$
values  and extrapolate the resultant series of $T_c(\nu)$ towards
$\nu \to 0$ using the leading order form
$T_c(\nu)/\varepsilon_F(\nu) =T_c/\varepsilon_F -
\mathrm{const}\cdot \nu^{1/3}$. This functional form is expected
from the analysis of the difference between the scattering
$T$-matrices on the lattice and in the continuum
\cite{inpreparation}.


\begin{figure}
\includegraphics[width=0.99\columnwidth,keepaspectratio=true]{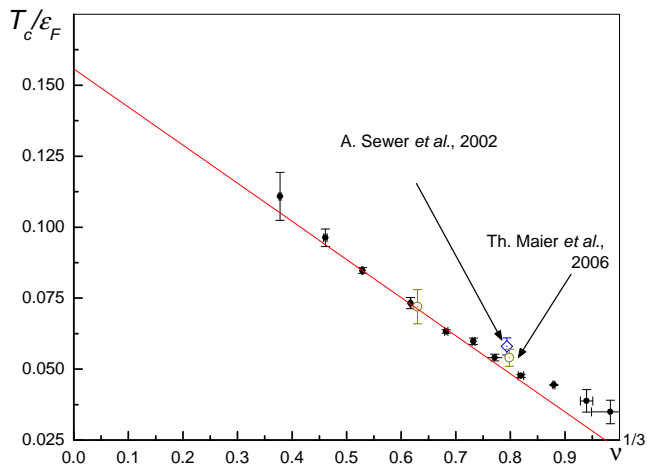}
\caption{The scaling of the lattice critical temperature with
filling factor (circles). The errorbars are one standard
deviation. The results of Ref.\ \cite{Zotos,DCA} at quarter
filling are also shown for a comparison. See the text for
discussion.}
 \label{fig-tc}
\end{figure}

Shown in Fig.\ \ref{fig-tc} are the simulation results for the
critical temperature at filling factors ranging from $0.95$ down
to $0.06$. We use system sizes up to $16^3$ sites with up to 300
fermions. It is clearly seen that starting from $\nu \approx 0.5$
the expected $\nu^{1/3}$ scaling holds very well and the
subleading corrections are negligible. On the other hand, close to
half-filling, $T_c(\nu)$ is essentially constant, (see,
\textit{e.g.} \cite{Micnas}).

Figure~\ref{fig-tc} shows a strong dependence of $T_c(\nu)$ on
$\nu$. This is in apparent contradiction with Ref.\ \cite{Bulgac}
which assumes no such dependence. This might be due to different
single-particle spectra $\epsilon_k\,$: Ref.\ \cite{Bulgac}
employs a parabolic spectrum with a spherically symmetric cutoff,
while we use a tight-binding spectrum over all of the Brillouin
zone. Our preliminary tests show that $\sim \nu^{1/3}$ corrections
do depend on the specific choice of single-particle spectrum, and
may even have different signs for different $\epsilon_k$.

The critical temperature we derive from Fig. \ref{fig-tc} is
$T_c=0.152(7) \varepsilon_F$. Various approximate schemes have in
the past yielded $T_c$ to be either above
\cite{mf,Holland-Timmermans, Perali} or below
\cite{LiuHu,Ohashi-Griffin,Haussmann94} the BEC limit
$T_\mathrm{BEC}=0.218 \varepsilon_F$. Our results clearly show
that it is below.

Previous numerical results were also in disagreement on whether
$T_c$ is higher or lower than $T_\mathrm{BEC}$: Ref.\
\cite{Wingate} quotes $T_c / \varepsilon_F =0.05$, but the
scattering length has not been determined precisely. Most
probably, this result corresponds to a deep BCS regime, where the
critical temperature is exponentially suppressed. Lee and
Sch\"{a}fer \cite{Lee-Schaefer} claim an upper limit $T_c < 0.14
\varepsilon_F$. This result is based on a study of the caloric
curve of a unitary Fermi gas down to $T/\varepsilon_F=0.14$ for
filling factors down to $\nu=0.5$. The caloric curve of Ref.\
\cite{Lee-Schaefer} shows no signs of divergent heat capacity
which would signal the phase transition. We find it not surprising
since at quarter filling $T_c(\nu=0.5)/\varepsilon_F \approx
0.05$, see Fig.\ \ref{fig-tc}. The simulations of Ref.\
\cite{Bulgac}, which are also based on a caloric curve study,
yield $T_c = 0.23(2) \varepsilon_F$. What this otherwise excellent
treatment lacks is an accurate finite-size and finite-density
analysis of the MC data.

The value of $T_c$ determined in this work cannot be directly
compared to the experimental result $T_c = 0.27(2) \varepsilon_F$
\cite{Thomas_expt} for a number of reasons. First, there are
strong indications that a presence of a trap significantly
enhances the transition temperature, see \textit{e.g.} Ref.\
\cite{Perali}. Second, the data analysis of Ref.\
\cite{Thomas_expt} relies on a mean-field approximate theory for
relating the empirical and actual temperature scales. In this
regard, it would be extremely interesting to see to what extent
the results of Ref.\ \cite{Thomas_expt} would be affected if a
different theoretical scheme is employed for thermometry.


\begin{figure}[htb]
\includegraphics[width=0.99\columnwidth,keepaspectratio=true]{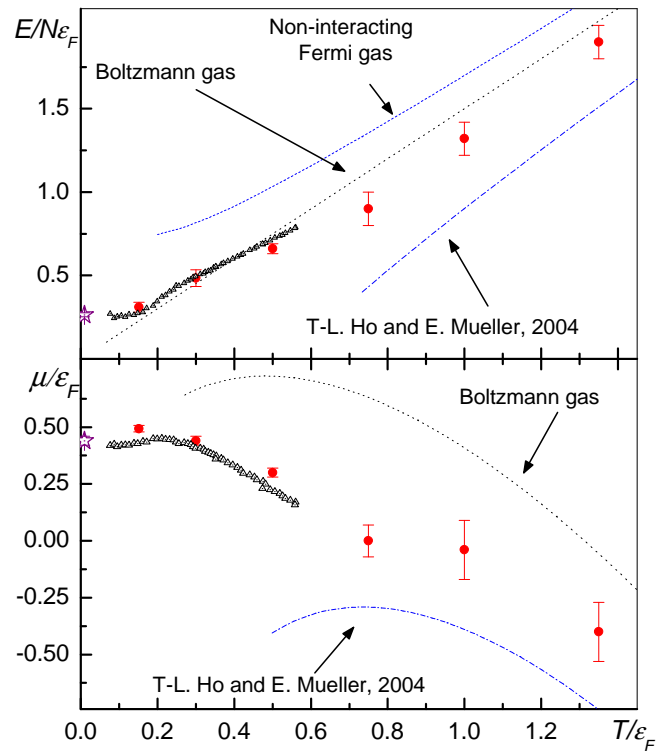}
\caption{The temperature dependence of the energy per particle
(upper panel) and chemical potential (lower panel) of the unitary
Fermi gas. Red circles are the MC results, black dotted lines and
blue dashed lines correspond to the Boltzmann and non-interacting
Fermi gases, respectively, the dot-fashed lines are the asymptotic
prediction of Ref.\ \cite{HoMueller} (plus the first virial Fermi
correction), black triangles are the path integral MC results of
Ref.\ \cite{Bulgac}, and the purple stars denote the ground-state
fixed-node MC results \cite{Giorgini-Carlson}. }
 \label{emuplot}
\end{figure}

At unitarity, the thermodynamic functions acquire a self-similar
form \cite{Ho}. For example, for the free energy one has:
\begin{equation} F(T,V,N) = f^{(F)}(T/\varepsilon_F) N
\varepsilon_F,
\label{F}
\end{equation}
where $N$ is the number of particles, $V$ is the volume, and
$f^{(F)}(x)$ is a dimensionless function.  Eq.\ (\ref{F}) allows
one to express all thermodynamic potentials in terms of energy per
particle  $f^{(E)}=E/N\varepsilon_F$ and rescaled chemical
potential $f^{(\mu)}=\mu/\varepsilon_F$. The latter quantities are
directly measurable numerically.

%

An analysis similar to the calculation of $T_c$  yields
\begin{align}
&E / (N\varepsilon_F) = 0.31(1), \\
&\mu / \varepsilon_F  = 0.493(14).
\label{magic_numbers}
\end{align}
For the pressure $P$ and entropy $S$, one than has
\begin{align}
&P / ( n\varepsilon_F) = 0.207(7), \\
&S / N = 0.16(2),
\label{magic_numbers_1}
\end{align}
which follows from (\ref{magic_numbers}) and exact relations
$PV=(2/3)E$ and $S = ( 5E/3 - \mu N )/NT$. Eqs.\
(\ref{magic_numbers})-(\ref{magic_numbers_1}) are for $T=T_c$.

Shown in Fig.\ \ref{emuplot} are our results for the dependence of
the energy per particle and chemical potential on temperature. In
the high-temperature simulations we use system sizes of up to
$32^3$ sites with up to 80 fermions.

As can be seen in Fig.\ \ref{emuplot}, our results for both
energy and chemical potential approach values close to the
fixed-node MC values \cite{Giorgini-Carlson} as $T \to 0$. For
$T/\varepsilon_F \leqslant 0.5$ our results are not far from the
curve of Ref.\ \cite{Bulgac}. As $T/\varepsilon_F \to \infty$,
both energy and chemical potential approach the virial expansion
\cite{HoMueller}  at high temperatures.


In conclusion, we have performed a determinant diagrammatic MC
simulations of a unitary Fermi gas by means of diluting the
attractive Hubbard model. In order to extract the continuum gas
behaviour we carefully treat both finite-size and lattice
corrections. We have determined the critical temperature
$T_c/\varepsilon_F = 0.152(7)$, the values of the thermodynamic
functions at criticality, and the overall shape of the
thermodynamic potentials from zero- to high-temperature regimes.

\acknowledgments{We thank A.~Bulgac, P.~Magierski and J. Drut for
providing us with their data. This research was enabled by
computational resources of the Center for Computational Sciences
and in part supported by the Laboratory Research and Development
program at Oak Ridge National Laboratory. Part of the simulations
were performed on the ``Hreidar'' cluster of ETH Z{\"u}rich. We
also acknowledge partial support by NSF grants Nos. PHY-0426881
and PHY-0456261 and by the Swiss National Science Foundation. }


\begin{figure*}[p]
\includegraphics[width=2\columnwidth]{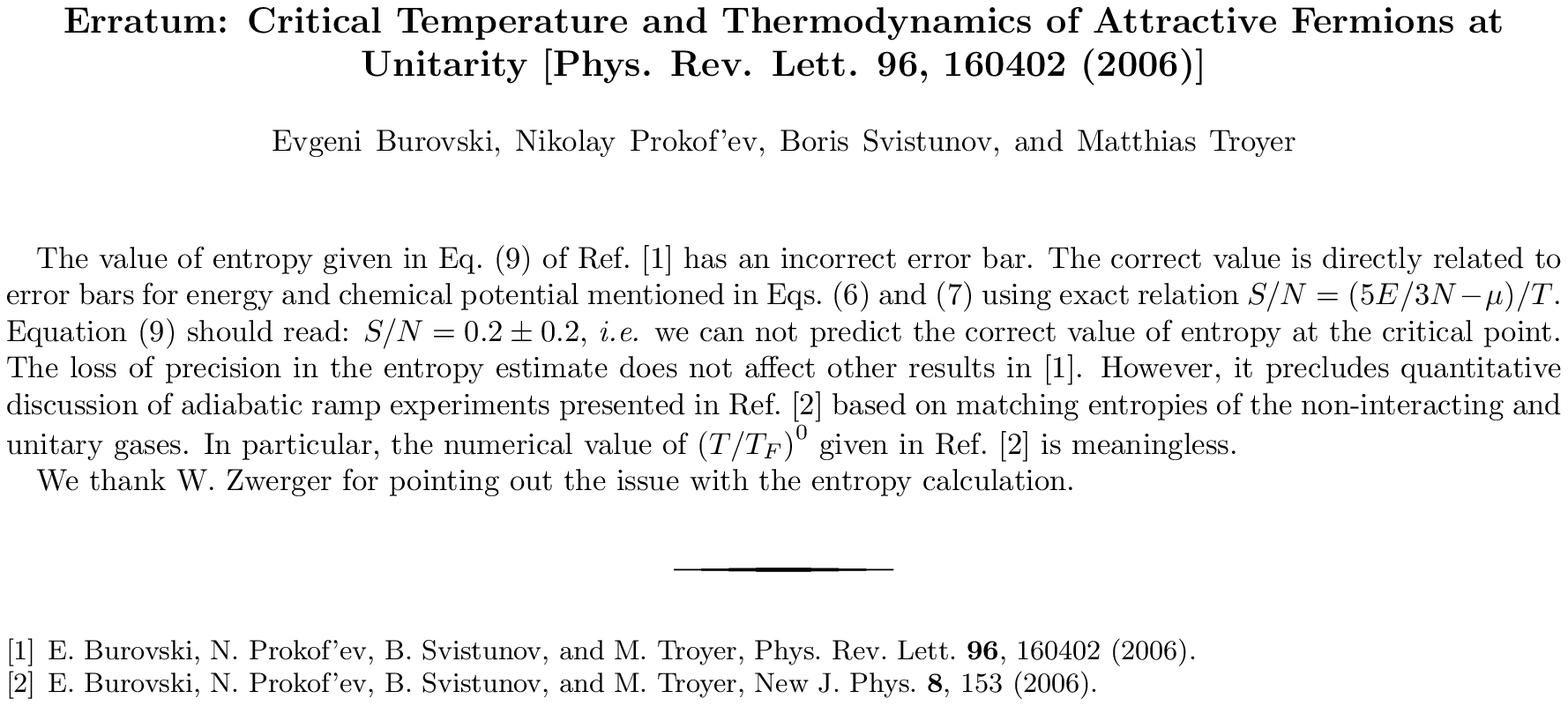}
\end{figure*}


\end{document}